\documentclass[12pt]{article}
\usepackage[utf8]{inputenc}
\usepackage{geometry}
\usepackage{amsmath,amssymb,amsthm,amstext}
\usepackage{mathtools}
\usepackage{siunitx}   
\usepackage{braket}
\usepackage{verbatim}
\usepackage{multirow}
\usepackage{dsfont}
\usepackage{siunitx}
\usepackage{float}
\usepackage{cite} 
\usepackage{graphicx,subfigure}
\usepackage{changebar}
\usepackage[clean]{revdiff}
\usepackage{hyperref}
\usepackage[colorinlistoftodos]{todonotes}
\definecolor{MyOrange}{rgb}{0.93, 0.53, 0.18}

\graphicspath{{figs/}}

\title{Dephasing versus collapse: Lessons from the tight-binding model with noise}
\author{Marco Hofmann and Barbara Drossel\\
\rchange{TU}{Technische Universität} Darmstadt, Institute for Condensed Matter Physics\\ Hochschulstrasse 6, 64289 Darmstadt, Germany}

\begin{document}
\maketitle
\begin{abstract}
Condensed matter physics at room temperature usually assumes that electrons in conductors can be described as spatially narrow wave packets - in contrast to what the Schrödinger equation would predict. How a finite-temperature environment can localize wave functions is still being debated. Here, we represent the environment by a fluctuating potential and investigate different unravellings of the Lindblad equation that describes the one-dimensional tight-binding model in the presence of such a potential. While all unravellings show a fast loss of phase coherence, only part of them lead to narrow wave packets, among them the quantum-state diffusion unravelling. Surprisingly, the decrease of the wave packet width for the quantum state diffusion model with increasing noise strength is slower than that of the phase coherence length. In addition to presenting analytical and numerical results, we also provide phenomenological explanations for them. We conclude that as long as no feedback between the wave function and the environment is taken into account, there will be no unique description of an open quantum system in terms of wave functions. We consider this to be an obstacle to understanding the quantum-classical transition.
\end{abstract}

\section{Introduction}

The transition from quantum to classical physics is one of the big unsolved puzzles in physics. One important feature of classical physics is the localization of objects. A macroscopic or mesoscopic object does not occur in superimposed states, except if temperatures are extremely low (as in quantum optomechanical experiments \cite{ockeloen2018stabilized}) or a system is prevented from interaction with the environment by some other means (as in double-slit experiments with bucky balls \cite{arndt1999wave}). The theoretical physics methods employed when dealing with finite-temperature many-particle systems reflect this assumption of localization \cite{drossel2019condensed}: Ab inito molecular dynamics simulations presume that the atoms or nuclei of the system are localized within a distance of the order of the thermal wavelength \cite{marx2000ab}; density functional theory starts from the assumption that nuclei have classical positions and are not entangled with each other \cite{kohn1999}; quantum-mechanical theories for the electric resistance of metals model electrons as wave packets with a limited spatial extension \cite{solyom2007fundamentals2}. With the exception of macroscopic quantum states such as superconductivity \cite{BCS}, we usually assume that the positions of atoms that constitute a solid or fluid and of the electrons that transport electric currents have at each moment in time a pretty well specified position. 

Localization of quantum particles happens also during a quantum measurement. Detectors such as photographic plates, photodiodes, Geiger counters, or CCDs are all based on the principle that the particle to be detected deposits energy at a specified position in the detector. This interaction then becomes amplified in order to generate a visible signal. Detectors are classical devices that have a macroscopic number of degrees of freedom and a finite temperature. 

The vast number of thermal degrees of freedom of the environment is considered crucial for inducing localization, as pointed out  by decoherence theory \cite{zurek2006decoherence}. This theory takes a reductionist approach and describes the degrees of freedom of the measurement apparatus by a many-particle Schr\"odinger equation, including an interaction with the quantum particle to be measured. Time evolution leads according to this description to the entanglement between the system and environment.  
Decoherence theory has been very successful at explaining why the reduced density matrix of the quantum system of interest becomes diagonal in the eigenbasis of the observable that is singled out by the interaction (which is the position for the situations we are considering here). However, according to various authors, decoherence theory cannot solve the puzzle why in a single run of a measurement only one outcome is observed  \cite{adler2003decoherence,schlosshauer2005decoherence,drossel2018contextual}, or why our observations of objects yield only one of the different possible positions that result from decoherence calculations. Additionally, the postulated immense amount of entanglement between a system and a finite-temperature bath consisting of a macroscopic number of degrees of freedom is inaccessible to any experimental test and can therefore be questioned.
 
The probably best-known proponent of the view that condensed matter theory cannot be reduced to a many-particle Schr\"odinger equation is  Anthony Leggett \cite{leggett1992nature}. His arguments are mainly based on what happens during a quantum measurement. According to Leggett, quantum mechanics must eventually break down in condensed matter systems as system size increases. 
Due to the mentioned problems of decoherence theory, various approaches to the interpretation of quantum mechanics place limits on the validity of unitary time evolution, as is done in particular in collapse models \cite{bassi2013models}. 

The idea that finite-temperature systems behave classically at sufficiently large scales led various earlier authors to describe the heat bath in a simplified manner as a classical fluctuating potential \cite{girvin1979exact}. A useful example system that describes an electron in a thermally fluctuating environment is the noisy (one-band) tight-binding model, which is essentially a discretised free Schrödinger equation that includes a white-noise potential. Different versions of such models have been studied in the past decades. If the noise is quenched, i.e., constant in time, one obtains the Anderson model which shows the famous phenomenon of Anderson localization \cite{anderson1958absence,tiggelen}. The eigenfunctions of the Hamiltonian are localized in one- and two-dimensional systems. In higher dimensions, they are localized if their energy is sufficiently far away from the band centre, or if randomness is strong enough. Fluctuating spatiotemporal potentials, in contrast, give delocalized wave functions as the width of the wave function increases with the square root of time \cite{ovchinnikov1974,Bouchaud1992,Saul1992,kang2009diffusion}. This means that classical fluctuating potentials are not an appropriate means to obtain the desired description of electrons as wave packets of a limited width.

However, there are other, empirically equivalent ways to describe the same system. As an alternative to modelling a stochastic system as an individual wave function trajectory, there is the description via the density matrix, which represents an ensemble of identically prepared quantum systems. In fact, since thermal noise can neither be  controlled nor measured on the microscopic level, the density matrix captures everything that is relevant for the outcome of any type of measurement that can be performed on the quantum particle.

The most general description of the time evolution of a density matrix in the absence of memory effects is a Lindblad equation \cite{breuer2002theory}. Many researchers in the field of open quantum systems interpret the Lindblad equation as an equation for the reduced density matrix of a system that is entangled with its environment, with the two together obeying unitary time evolution. But since the Lindblad equation can be derived from just a few logical requirements (in particular the Markov property), it can as well be interpreted as describing a system in an environment that is not a quantum system with unitary time evolution, but a classical environment. Specifically, the ensemble description of the time evolution of a quantum particle in a fluctuating classical potential is also given by a Lindblad equation.

Now, a Lindblad equation can be unravelled in different ways into the stochastic time evolution of an ensemble of wave functions. The original time evolution with a classical stochastic potential is only one possibility. Two examples of other unravellings are jump dynamics \cite{ghirardi1986unified} and  quantum state diffusion \cite{gisin1992quantum}, both of which lead to wave functions that retain a limited width even at large times. There is no way to discriminate by experiment between the different unravellings. This means that the same model that gives rise to the broadening wave function in a fluctuating potential can also generate wave function localization if represented in a different, but empirically equivalent way. The publications that have so far studied the noisy tight-binding model have not investigated wave function dynamics obtained from these other, empirically equivalent unravellings. However, apart from leading to the intuitively satisfying picture of electrons in conductors as wave packets, these unravellings yield a wealth of additional insights. 

It is the purpose of this paper to explore the different types of unravellings of the Lindblad equation that is equivalent to the white-noise tight-binding model. While the original model gives an ever broadening wave function that shows dephasing between its values at sites that are further apart than the phase coherence length, the collapse models show a rapid transition to a narrow wave packet that performs diffusive motion. 
We perform computer simulations of the different unravellings and evaluate ensemble-averaged quantities, such as centre-of-mass motion and wave packet width. By combining these simulations with analytical calculations and phenomenological arguments, we bring together various known results spread over different papers, provide a deeper understanding of these results, and reveal new properties that were hitherto not known. We will also discuss how the features of the different unravellings are connected via the same ensemble description.

In the discussion at the end of the paper, we will address the conceptual questions raised by the investigation:  What is the 'real' unravelling, if they are all empirically equivalent? What must be changed in the modelling of the heat bath and of its interaction with electrons to obtain spatially narrow wave packets as the unique solution?

\section{Model}

In this section, we first present the model of an electron on a lattice in a fluctuating potential. We then derive the corresponding Lindblad equation, and from it the non-linear stochastic wave function evolutions that correspond to the two most common unravellings, which are quantum-state diffusion and jump dynamics. While the dynamics in the fluctuating potential and the corresponding Lindblad equation are known from the literature, the two non-linear unravellings have not yet been studied for this model.

We consider non-interacting electrons in a one-dimensional lattice in the tight-binding model and add white noise to it, modelled as a stochastic potential. The noise at different lattice sites is uncorrelated and is meant to describe the influence of thermal fluctuations. This model is the simplest possible model for electrons in the conduction band of a metal at finite temperature. 

The tight-binding model uses Wannier functions as a basis. They are essentially Fourier transforms of the Bloch functions and are localized in the vicinity of lattice sites when the binding energy dominates over the kinetic energy. 
Written in the Wannier basis $\{\ket{n}\}$, the state of an electron $\ket{\psi}$ is given by $\ket{\psi}=\sum_n c_n \ket{n}$, where $c_n$ is the amplitude of the wave function at site $n$. The time evolution of the electron state is given by 
\begin{eqnarray}\label{eq:WNP}
\ket{d\psi}&=& -\frac{i}{\hbar}\hat{\tilde H}\ket{\psi}d\tau -\frac{i}{\hbar}\sqrt{\Gamma}\sum_n \ket{n}\braket{n|\psi}d\tilde{W}_n - \frac{\Gamma}{2\hbar^2}\ket{\psi}d\tau\nonumber\\
&=&
 \frac{i\hbar}{2ma^2} \sum_n (\ket{n}\bra{n+1} + \ket{n+1}\bra{n} - 2 \ket{n}\bra{n} )\ket{\psi} d\tau\\
&& -i\frac{\sqrt{\Gamma}}{\hbar}\sum_n \ket{n}\braket{n|\psi}d\tilde{W}_n - \frac{\Gamma}{2\hbar^2}\ket{\psi}d\tau \, ,\nonumber
\end{eqnarray}
or, if written in terms of the coefficients $c_n$, 
\begin{equation}\label{eq:WNPcoeff_old}
dc_n = \frac{i\hbar}{2ma^2}(c_{n+1}+c_{n-1}-2c_n) d\tau -\frac{i}{\hbar}\sqrt{\Gamma} c_n d\tilde{W}_n - \frac{\Gamma}{2\hbar^2}c_n d\tau\, .
\end{equation} 
The first term on the right-hand side is the kinetic energy term. The simplest way to derive it is not via the tight-binding approximation, but by discretising the Schrödinger equation. The lattice constant is denoted as $a$. The second term is the stochastically fluctuating potential, and it replaces the classical time-constant potential that usually occurs in the Schrödinger equation. 
The stochastic potential $\tilde{W}_n(\tau)$ at site $n$ is modelled as a \textit{Wiener process}, which is uncorrelated in time and has increments chosen from a Gaussian distribution
	\begin{equation} 
	p(d\tilde{W})=\frac{1}{\sqrt{2\pi d\tau}} e^{-\frac{d\tilde{W}^2}{2d\tau}}.
\end{equation}
The last term must be added in order to preserve normalization of the wave function. The notation in \eqref{eq:WNP} and \eqref{eq:WNPcoeff_old} means that we are following the Ito calculus.

The change of the norm of the wave function is given by
\begin{eqnarray}\label{eq:calcnorm}
    d\sum_n c_n^*c_n &=& \sum_n c_n^* dc_n + \sum_n c_n dc_n^*  + \sum_n dc_n^* dc_n\nonumber\\
        &=& \frac{\Gamma}{\hbar^2}\sum_n |c_n|^2 (d\tilde W_n^2 - d\tau) = 0\, .
\end{eqnarray}
In the last step, we have used $d\tilde W_n^2 = d\tau$ \cite{Jacobs2010}. The random deviations of $d\tilde W_n^2 $ from $d\tau$ are of higher order in $d\tau$ and vanish therefore in the limit of continuous time. Calculation \eqref{eq:calcnorm} shows that we need the last term in equation \eqref{eq:WNPcoeff_old} in order to preserve normalization of the wave function. Without the last term, the weights $|c_n|^2$ would increase on average due to the Wiener process. Other ways of normalizing the wave function are mathematically equivalent to this one, as the differences would be of higher than linear order in $d\tau$. The need to normalize this model explicitly was discussed in the 1990s due to artifacts occurring in computer simulations \cite{Feng1990,Medina1991,Bouchaud1992}, but these authors did not derive the simple form \eqref{eq:WNPcoeff_old} chosen by us.

On a lattice with a lattice constant $a$, the energy remains limited due to the maximum possible wave number $\pi/a$. This limitation is essential if the model shall describe a particle in a finite-temperature environment. 
In contrast, models with fluctuating potentials in continuous space can cause the expectation value of the momentum to perform a random walk \cite{Jayannavar1982}, the energy of which is unbounded. That result was found with the potential being delta-correlated in time and having a finite correlation length in space. However, depending on the spatial and temporal correlations of the fluctuating potential, a slower increase of energy with time \cite{Lebedev1995} and even a constant energy can also occur in a spatially continuous system \cite{heinrichs1992diffusion}. The latter result was obtained when noise was delta correlated both in space and time. 

To simplify notation, we will switch for the remainder of this paper to dimensionless units and define $$t \coloneqq \frac{\hbar}{2ma^2} \tau\, , \quad \gamma\coloneqq \frac{2ma^2\Gamma}{\hbar^3}\, , \quad d W_n \coloneqq \sqrt{\frac{\hbar}{2ma^2}} d\tilde W_n \, ,\quad \hat{H} \coloneqq \frac{2ma^2}{\hbar^2} \hat {\tilde H}\, .$$ 
This ensures  that $\frac{dW^2}{dt} = \frac{d\tilde W^2}{d\tau}$.
Equation \eqref{eq:WNPcoeff_old} then becomes
\begin{equation} \label{eq:WNPcoeff}
dc_n = i(c_{n+1}+c_{n-1}-2c_n) dt -i\sqrt{\gamma} c_n dW_n - \frac{\gamma}{2}c_n dt\, .
\end{equation}

\subsection{The Lindblad equation}
The Lindblad equation for our model is the equation for a statistical ensemble of systems that evolve according to \eqref{eq:WNPcoeff}. Since the Lindblad equation is linear in the density matrix, it is sufficient to evaluate the change during $dt$ of an initially pure state,
\begin{eqnarray}
d(\braket{n|\psi}\braket{\psi|m})&=& \braket{n|d\psi}\braket{\psi|m}+ \braket{n|\psi}\braket{d\psi|m} + \braket{n|d\psi}\braket{d\psi|m}\nonumber\\
&=& i (\braket{n+1|\psi}\braket{\psi|m} + \braket{n-1|\psi}\braket{\psi|m} \nonumber\\
&&- \braket{n|\psi}\braket{\psi|m+1}-\braket{n|\psi}\braket{\psi|m-1})dt \nonumber\\
&& + \gamma\braket{n|\psi}\braket{\psi|m}dW_n dW_m -\gamma\braket{n|\psi}\braket{\psi|m}dt. \nonumber
\end{eqnarray}

Throughout the paper, we denote the mean over realizations of the noise as $\mathbf{M}$, which we refer to as the ensemble mean, and denote the quantum mechanical expectation value as $\langle \dots\rangle$. Taking the ensemble mean in the equation above gives the elements of the density matrix
\begin{eqnarray} \label{eq:Lindbladnm}
 d\rho_{n,m}&=& d\mathbf{M}[\braket{n|\psi}\braket{\psi|m}]\nonumber\\
 &=& i(\rho_{n+1,m}+\rho_{n-1,m}-\rho_{n,m+1}-\rho_{n,m-1})dt\nonumber\\
&& + \gamma\rho_{n,m}\mathbf{M}[dW_n dW_m] -\gamma\rho_{n,m}dt\nonumber\\
&=& i(\rho_{n+1,m}+\rho_{n-1,m}-\rho_{n,m+1}-\rho_{n,m-1})dt\nonumber\\
 &&+ \gamma\rho_{n,m}\delta_{nm}dt -\gamma\rho_{n,m}dt.
\end{eqnarray}
In matrix notation the Lindblad equation of our model becomes
\begin{equation}\label{eq:ULindblad}
\dot{\rho} = -i[\hat{H},\rho] +\gamma(diag[\rho]-\rho)
\end{equation}
where $diag[\rho]$ is the diagonal of $\rho$ with all other elements being zero. This is the discrete version of the Lindblad equation given e.g. in Ref. \cite{chechetkin1987quantum} for the continuous analogue of \eqref{eq:WNPcoeff_old}. Our discrete equation \eqref{eq:ULindblad} was obtained in a different context by Kenkre and Brown~\cite{KenkreStochasticLiouville}.
The first term is the von-Neumann term of the Hamiltonian that couples adjacent lattice sites, the second term is the effect of the noise and the third term preserves the trace of the density matrix. To write equation \eqref{eq:ULindblad} in the familiar Lindblad form, we insert the identities $\mathds{1}=\sum_n \ket{n}\bra{n}$ and $1=\braket{n|n}$ at the appropriate places:
\begin{eqnarray}
\dot{\rho}&=& -i[\hat{H},\rho] +\sum_n \gamma\left(\ket{n}\rho_{nn} \bra{n} - \frac{1}{2}\ket{n}\bra{n}\rho - \frac{1}{2}\rho\ket{n}\bra{n}\right)\nonumber\\
&=& -i[\hat{H},\rho] +\sum_n \gamma \left(\ket{n}\bra{n}\rho \ket{n}\bra{n} - \frac{1}{2}\ket{n}\braket{n|n}\bra{n}\rho - \frac{1}{2}\rho\ket{n}\braket{n|n}\bra{n}\right)\nonumber\\
&=&  -i[\hat{H},\rho] + \sum_n \gamma \left(A_n \rho A_n^\dagger -\frac{1}{2}A_n^\dagger A_n \rho - \frac{1}{2}\rho A_n^\dagger A_n\right) \label{lindbladAn}
\end{eqnarray}
The Lindblad operators of our model are the projectors $A_n=A_n^\dagger=\ket{n}\bra{n}$ onto the lattice sites $n$. The rates $\gamma_n = \gamma$ are independent of $n$ and are given by the noise strength $\gamma$. Lindblad equations are the most general linear  Markovian time evolution equations for the density matrix. Linearity permits the ensemble interpretation of the density matrix. Furthermore, the equation preserves the trace, self-adjointness and positivity of the density matrix.

\subsection{Unravelling via  quantum state diffusion}

Now there are several different ways to 'unravel' a Lindblad equation in terms of stochastic wave function dynamics such that the ensemble average of the time evolution is described by the Lindblad equation. Our starting equation \eqref{eq:WNPcoeff} is one possible unravelling of the Lindblad equation \eqref{eq:ULindblad}, and we call it the 'white-noise potential unravelling'. There is an infinity of different possible unravellings \cite{rigo1997continuous}, and we investigate the two most prominent ones, which are  quantum state diffusion and the quantum jump model. As long as there is no possibility to track, control,  or time-reverse the noise there is no way to distinguish empirically between the different unravellings since all observable results depend only on the density matrix. If it was possible to track the noise, one could in principle run the identical time evolution many times and measure the full wave function by using quantum state tomography \cite{d2001quantum}.

Gisin and Percival \cite{Gisin1992} found that every Lindblad equation can be unravelled into a diffusion equation of the form \begin{equation}
\ket{d\psi}= \ket{v}dt + \sum_n \ket{u_n}d\xi_n. 
\end{equation} 
The first term is a drift term and the second term is a stochastic diffusion term with complex Brownian motion $d\xi_n=\text{Re}[d\xi_{n}]+i \text{Im}[d\xi_{n}]$. The real and imaginary part of $d\xi$ are uncorrelated real-valued Wiener processes with variance $dt/2$, i.e. they fulfil the following relations:
\begin{eqnarray}
\mathbf{M}(\text{Re}[d\xi_{n}]\text{Im}[d\xi_{m}])&=& 0\nonumber\\ 
\mathbf{M}(\text{Re}[d\xi_{n}]\text{Re}[d\xi_{m}])=
\mathbf{M}(\text{Im}[d\xi_{n}]\text{Im}[d\xi_{m}])&=&
\frac{dt}{2}\delta_{nm}\\
\mathbf{M}(d\xi_n d\xi_m^\ast)&=& dt \delta_{nm}.\nonumber
\end{eqnarray}
Here, $\mathbf{M}$ denotes again the ensemble mean over different realizations of the noise. 
From the requirements that the wave function is normalized and that the time evolution of an ensemble satisfies the Lindblad equation, it follows that \cite{Gisin1992}
\begin{eqnarray}
\ket{v}&=&-i\hat{H}\ket{\psi} + \sum_n \gamma_n \left(
\langle A_n^\dagger \rangle A_n - \frac{1}{2}A_n^\dagger A_n - \frac{1}{2} \langle A_n^\dagger \rangle \langle A_n \rangle
\right)\ket{\psi}\,\,\text{and}\nonumber
\\
\ket{u_n}&=& \sum_n \sqrt{\gamma_n} (A_n-\langle A_n\rangle)\ket{\psi}\, .
\end{eqnarray}
This is the general form of the quantum state diffusion unravelling for a Lindblad equation with Hamiltonian $\hat H$ and Lindblad operators $A_n$. The expressions $\langle A_n \rangle$ are the short notation for the quantum mechanical expectation values $\langle\psi|A_n|\psi\rangle$.
For our specific model \eqref{eq:ULindblad} the Lindblad operators are projectors onto lattice sites, $A_n^\dagger = A_n =\ket{n}\bra{n}$, and $\gamma_n=\gamma$ for all $n$. The quantum state diffusion unravelling becomes then
 \begin{eqnarray}
 \ket{d\psi}&=&-i\hat{H}\ket{\psi}dt -\frac{\gamma}{2} \sum_n (\ket{n}\bra{n}-\vert c_n \vert^2)^2
 \ket{\psi} dt\nonumber\\
 &&+ \sqrt{\gamma}\sum_n (\ket{n}\bra{n}-\vert c_n \vert^2)\ket{\psi}d\xi_n\, , \label{eq:UQSD}
 \end{eqnarray}
 which is equivalent to
 \begin{eqnarray}
 \mathrm{d}c_n&=&i(c_{n+1}+c_{n-1}-2c_n)\mathrm{d}t+\gamma\left(\vert c_n \vert^2-\frac{1}{2}-\frac{1}{2}\sum_m \vert c_m\vert ^4\right)c_n\mathrm{d}t\nonumber\\
 &&+\sqrt{\gamma}\left(\mathrm{d}\xi_n-\sum_m \vert c_m \vert^2 \mathrm{d}\xi_m\right)c_n. \label{eq:UQSDcoeff}
 \end{eqnarray}
 
 The first term is the kinetic energy term that causes a broadening of the wave packet, which is countered by the other two terms. The first contribution to the second term is a non-linear growth term. Its effect is that the weight $|c_n|^2$ on sites that already have a large weight increases most. The other two contributions to the second term cause a decrease of all amplitudes, so that normalization of the wave function is preserved. The third term is stochastic. The absolute value of the expression in the bracket is smallest when $|c_n|^2$ is largest, meaning that the wave function fluctuates less on sites where its weight is larger. In the absence of the kinetic energy term, the stationary solutions of \eqref{eq:UQSD} are position eigenstates. Together with the kinetic energy term, the time evolution according to \eqref{eq:UQSD} results in wave packets the width of which depends on the strength of the noise $\gamma$. For large $\gamma$ the Lindblad term dominates dynamics, and we can expect the wave function to be very narrow. In fact,  quantum state diffusion has been shown to be the unravelling with the fastest rate of localization \cite{rigo1997continuous,percival1994localization}, which can be made even faster by using real instead of complex noise. \rold{(}However, the authors performed their calculations \rchange{in the limit of a wide-open quantum system, where the Hamiltonian term is neglected.)} {for a system in which the noise is strong compared to the Hamiltonian term and the latter is therefore neglected. Such a system is called a wide-open quantum system.}
 
 There exists a linear variant of the  quantum state diffusion unravelling, where the terms that are non-linear in the wave function are dropped \cite{Percival1998}. The wave functions resulting from this unravelling are however not normalized, leading to the dominance of one wave function in the ensemble after some time. Only if the noise is purely imaginary normalization is preserved - and the resulting equation for the wave function is identical to our 'white-noise potential unravelling' \eqref{eq:WNPcoeff}.
 
\subsection{Unravelling via quantum jumps}

The quantum jump unravelling was first introduced by Dalibard et al. \cite{Dalibard1992} and Carmichael \cite{Carmichael1993}. It is the most intuitive unravelling as it is a straightforward interpretation of the two types of terms in the Lindblad equation \eqref{lindbladAn}. The time evolution of a quantum state is a combination of a continuous evolution due to the Hamiltonian and to the outflow of probability, and of random transitions due to applying one of the Lindblad operators $A_n$. For our model, the quantum jump unravelling takes the simple form
\begin{eqnarray}\label{eq:UQJ}
\ket{\psi(t+\delta t)}=
\begin{cases}
\ket{\psi(t)} -i\hat{H}\ket{\psi(t)}\delta t & \text{ with probability } 1-\gamma \delta t  \\
\ket{n} & \text{ with probability } \vert c_n \vert^2 \gamma \delta t.
\end{cases}
\end{eqnarray}
With the rate $\gamma$, the quantum state collapses to one of the localized states $\ket{n}$, with the relative probabilities given by the values $\vert c_n \vert^2$ just before the collapse. The probability that no jump occurs in time $t$ is $\exp(-\gamma t)$. This means the waiting time $\tau$ between two consecutive jumps is exponentially distributed with mean $\gamma^{-1}$,
\begin{equation}\label{eq:taudistrib}
p(\tau)=\gamma e^{-\gamma\tau}.
\end{equation}
This model thus describes a particle that propagates freely on a lattice and whose position is measured with the rate $\gamma$.

\section{Results}

All three unravellings \eqref{eq:WNPcoeff}, \eqref{eq:UQSD}, and \eqref{eq:UQJ} give the same results when quantities are evaluated that are
 ensemble averages of physical observables  over different realizations of the noise. This is because the Lindblad equation \eqref{eq:ULindblad} is the same for all three unravellings, and all ensemble expectation values can be calculated with the Lindblad equation. However, it is in some cases more convenient to use one of the unravellings for a calculation. Usually, unravellings are used for numerical evaluations, but in the following we will use them also for analytical calculations that we will compare to computer simulations. All three unravellings and also the Lindblad equation will be used to obtain a variety of insights into the properties of our system.

The main results of this section are (i) a direct derivation of the diffusion constant that characterizes the increase of the variance of the position using the quantum jump unravelling (ii) a solution of the Lindblad equation in the limit of strong noise (iii) a phenomenological argument for the subdiffusive behavior of the centre of mass of the wave function in the fluctuating potential (iv) the derivation of a spatial coherence length (v) the discovery of an unexpected scaling behavior of the time evolution of the mean wave packet width in the quantum state diffusion model (vi) the time evolution of the participation number in the quantum state diffusion model.

\subsection{Linear increase of the  variance of the position}
It is a well-known result \cite{ovchinnikov1974,Bouchaud1992,Saul1992,kang2009diffusion} that the ensemble-averaged second moment of the position operator $\hat{x}=\sum_n n\ket{n}\bra{n}$ of our model increases linearly in time. If we choose the origin of the spatial coordinate such that $\mathbf{M}\langle\hat{x}\rangle=0$ at $t=0$, this means that 
\begin{equation}\label{eq:diffusion}
\mathbf{M}\langle\hat{x}^2\rangle= D\cdot t
\end{equation}
with a diffusion constant $D$. 
With the quantum jump unravelling \eqref{eq:UQJ}, the diffusive behavior of the squared distance from the initial position is completely obvious as the jump process is a random walk. This random walk is continuous in time with the distribution of waiting times given by \eqref{eq:taudistrib}, and with the distribution of jump distances being determined by the free spreading of the wave function between jumps. 

We show in the following how the quantum jump unravelling can be used to calculate the diffusion constant in a direct manner.
 The spreading of an initially fully localized wave packet on a one-dimensional lattice has recently been calculated by Schönhammer \cite{Hammer}, with the result that the variance of the wave packet increases in the absence of noise as
\begin{equation}\label{eq:fpvar}
\sigma_0^2(\hat{x})(t)=\langle\hat{x}^2\rangle_0(t)=2t^2\, .
\end{equation}
We denote with the index $0$ the absence of noise. 
Based on  \eqref{eq:fpvar} and \eqref{eq:taudistrib}, we calculate the mean squared jump distance
\begin{equation} \label{msjd}
\mathbf{M}[(\Delta x)^2] = \int_{0}^\infty dt \sigma_0^2(\hat{x})(t) p(t)= {2 \gamma}\int_{0}^\infty dt\, t^2 e^{-\gamma t} = \frac{4}{\gamma^2}\, .
\end{equation}
The mean squared distance $\mathbf{M}\langle\hat{x}^2\rangle$ covered by the jump process during time $t$ is obtained from $\mathbf{M}[(\Delta x)^2]$ by multiplying with the expected number of jumps $\gamma t$ during the time $t$. We therefore obtain the diffusion constant $D$
by multiplying the mean squared jump distance with the inverse  time $\gamma$ between jumps,
\begin{equation}\label{eq:D}
	D=\gamma\mathbf{M}[(\Delta x)^2]=\frac{4}{\gamma}.
\end{equation}
The noise strength $\gamma$ is in the denominator, implying that diffusion is slower when noise is stronger. This is a version of the quantum Zeno effect \cite{Zeno}: if the position is measured more often, it becomes less likely for the particle to jump to another lattice site.

The result \eqref{eq:D} for the diffusion constant has been calculated in \cite{ovchinnikov1974} and recently in~\cite{Hislop2019} by means of a Laplace transformation of the density matrix. Our derivation is a briefer and more intuitive alternative.
We use the result \eqref{eq:D} to check the accuracy of the numerical simulations for the different unravellings. All simulations in this paper were performed with the Euler-Maruyama method~\cite{Kloeden} with step size $\Delta t= 10^{-4}$ in dimensionless units. The initial state was chosen as a real-valued gaussian wave packet with finite width.
The quantum mechanical expectation of the squared position operator was calculated for various instants of time from the numerical solutions of \eqref{eq:WNPcoeff} and \eqref{eq:UQSDcoeff} as $\langle \hat{x}^2\rangle=\sum_n \vert c_n \vert^2 n^2$. The results were averaged over different realizations of the noise. The system size was chosen appropriately large (up to 1000 lattice sites) such that boundary effects could be neglected.
\begin{figure}
	\includegraphics[width=0.9\textwidth]{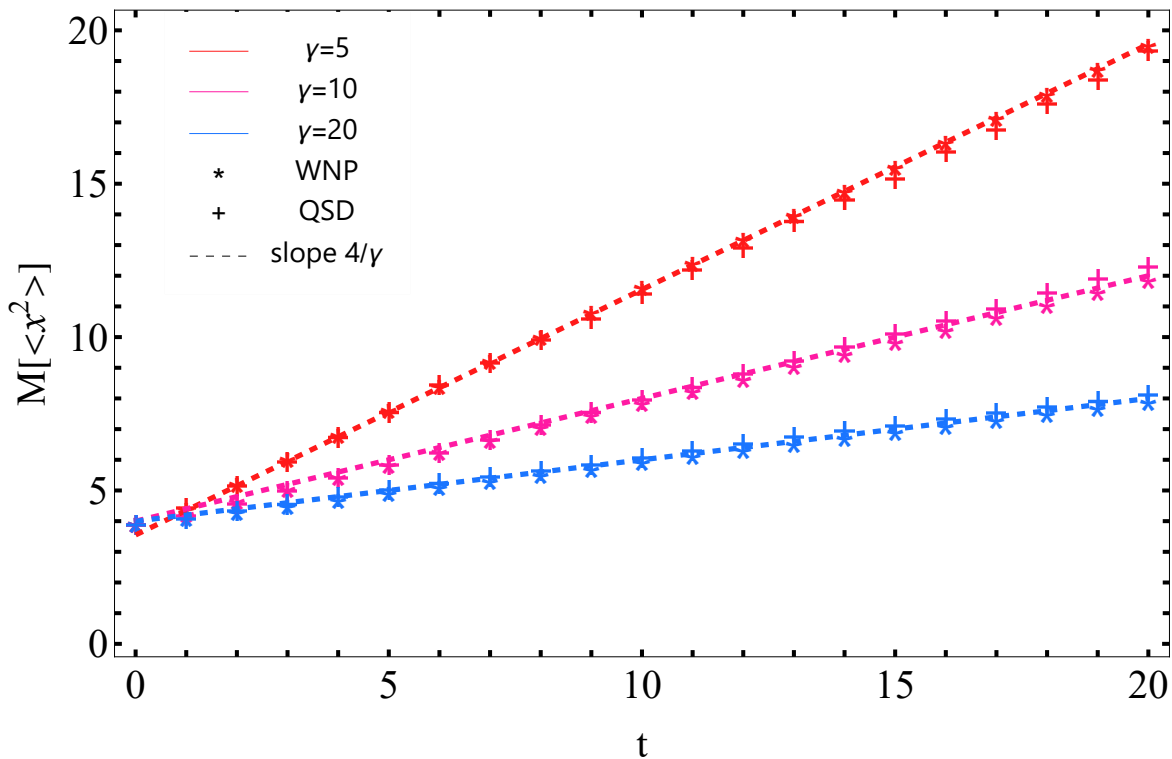}
	\caption{Ensemble mean of the second moment of the position operator over time. We compare numerical results (stars) of the white-noise potential (WNP) unravelling and the  quantum state diffusion (QSD) unravelling (crosses)  with the diffusive behavior at times $t  \gg 1/\gamma$ predicted by equation \eqref{eq:D} (dashed lines) for the noise strengths $\gamma=2ma^2\Gamma\hbar^{-3} =5,10,20$.
	The initial state is a gaussian wave packet with variance $\sigma^2 = 4$. The results are averaged over $10^4$ trajectories.
	}
	\label{fig:MSM}
\end{figure}
Figure~\ref{fig:MSM} compares the mean squared position from numerical simulations of the white-noise potential unravelling \eqref{eq:WNPcoeff} and the  quantum state diffusion unravelling \eqref{eq:UQSD} with the slope of the analytical solution \eqref{eq:D} which holds for $\gamma t \gg 1$ for different noise strengths. The agreement of the simulations with the analytical result is very good already for short times.

\rnew{Our simulation results shown in Figure \ref{fig:MSM} confirm that the diffusion equation \eqref{eq:diffusion} with diffusion constant \eqref{eq:D} is valid for all unravellings of the Lindblad equation \eqref{eq:ULindblad} for times $t\gg\gamma^{-1}$.} For unravellings \eqref{eq:UQSDcoeff} and \eqref{eq:UQJ} that lead to wave functions with a finite width, the diffusive increase of the second moment of the position operator implies that the centre of mass of the wave function also performs a diffusive motion. This is because the squared width of the wave packet is given by
\begin{equation}\label{sq}
\sigma^2(\hat{x})(t)=\langle \hat{x}^2 \rangle (t) - \langle \hat{x}\rangle ^2 (t).
\end{equation}
Since $\sigma^2(\hat{x})(t)$  remains finite for $t\to\infty$ for unravellings \eqref{eq:UQSDcoeff} and \eqref{eq:UQJ}, $\langle \hat{x}^2 \rangle (t)$ and $\langle \hat{x}\rangle ^2 (t)$ must increase in the same way with $t$ so that their difference remains finite.\\

\subsection{Time-scale separation between fast dephasing and slow diffusion in the limit of large noise} \label{sec:timescaleseparation}

For $\gamma \gg 1$, the phase fluctuations of the $c_n$ in \eqref{eq:WNPcoeff} caused by the stochastic potential become much faster than the kinetic energy term. This means that the phase correlations between the $c_n$ decay much faster than their amplitudes change. While phase correlations decay with a characteristic time $1/\gamma$, the time required for the weights $|c_n|^2$ to shift to neighbouring sites is proportional to $\gamma$, as indicated by the diffusion constant $D$ being proportional to $1/\gamma$, see~\eqref{eq:D}. The authors of \cite{amir2009classical} made use of this time-scale separation to show that the on-site probabilities $|c_n|^2$ obey a classical diffusion equation in this limit. We will in the following perform this time-scale separation in the Lindblad equation, which allows us to express the time evolution of all elements of the density matrix in terms of its diagonal elements only in the limit $\gamma \gg 1$.

The stationary solution of \eqref{eq:Lindbladnm} is 
\begin{equation}
    \rho_{n,m}^{st} = \frac 1 N \delta_{n,m}
    \end{equation}
 with $N$ being the number of lattice sites and assuming periodic boundary conditions. Relaxation to this stationary state involves a decay of the off-diagonal elements and a shift of the weights $\rho_{n,n}$ between lattice sites towards an equal distribution.

For an off-diagonal element $\rho_{n+k,n}$, the Lindblad equation \eqref{eq:Lindbladnm} becomes
\begin{equation}
    \dot{\rho}_{n+k,n}= i(\rho_{n+k+1,n}+\rho_{n+k-1,n}-\rho_{n+k,n+1}-\rho_{n+k,n-1}) - \gamma \rho_{n+k,n}\, .
\end{equation}
We choose w.l.o.g. $k>0$. The last term on the right-hand side, which is proportional to $\gamma$, dominates the dynamics, meaning that ${\rho}_{n+k,n}$ relaxes within a short time of the order $\gamma^{-1}$ to the value that makes the r.h.s. zero. This in turn means that ${\rho}_{n+k,n}$ is smaller by  a factor of the order of $\gamma^{-1}$ than the largest matrix element in the first term. Starting from the diagonal elements (for which $k=0$), we can thus conclude that the elements in the $k$th off-diagonal are smaller by a factor $\gamma^{-k}$ than the diagonal terms.  We use this finding in order to calculate the slow dynamics of the diagonal elements. The equations for the elements on the diagonal and the two neighbouring off-diagonals are
\begin{eqnarray}
    \dot{\rho}_{n,n}&=&i(\rho_{n+1,n}+\rho_{n-1,n}-\rho_{n,n+1}-\rho_{n,n-1})\label{eq:k}\\
   &=& -2\text{Im}[\rho_{n+1,n}+\rho_{n-1,n}]\nonumber\\
    \dot{\rho}_{n+1,n}&=& i(\rho_{n+2,n}+\rho_{n,n}-\rho_{n+1,n+1}-\rho_{n+1,n-1}) -  \gamma\rho_{n+1,n}\label{eq:k1}\\
   &\simeq& i(\rho_{n,n}-\rho_{n+1,n+1})-\gamma \rho_{n+1,n}\nonumber\\
     \dot{\rho}_{n-1,n}&=& i(\rho_{n,n}+\rho_{n-2,n}-\rho_{n-1,n+1}-\rho_{n-1,n-1}) - \gamma \rho_{n-1,n}\label{eq:k-1}\\
     &\simeq& i(\rho_{n,n}-\rho_{n-1,n-1})-\gamma \rho_{n-1,n}.\nonumber
     \end{eqnarray}
    In the last two equations, we kept only the leading terms. 
     For $\gamma t\gg 1$, we can exploit the time-scale separation due to the fast relaxation of the off-diagonal elements and set the left-hand sides of \eqref{eq:k1} and \eqref{eq:k-1} to zero. This gives
     \begin{eqnarray}
    \rho_{n+1,n}&\approx&\frac{i}{\gamma}(\rho_{n,n}-\rho_{n+1,n+1})\nonumber\\
    \rho_{n-1,n}&\approx&\frac{i}{\gamma}(\rho_{n,n}-\rho_{n-1,n-1})\, ,\label{eq:diffnnp1}\end{eqnarray}
which we insert into \eqref{eq:k} to obtain \cite{amir2009classical}
\begin{equation}\label{eq:diffnn}
    \dot{\rho}_{n,n}=\frac{2}{\gamma}(\rho_{n+1,n+1}+\rho_{n-1,n-1}-2\rho_{n,n})\, .
\end{equation}
This is the discretised  version of a diffusion equation $\dot p(x) = \frac D 2 \Delta p$ for the weights $p(x) = {\rho}_{n,n}$ with $D=4/\gamma$. The solution of this equation is $p \propto \exp[-x^2/2Dt]$ and gives an increase $\sigma_x^2 = Dt$ in the variance of the ensemble distribution of $x$.
This result is identical to our above result \eqref{eq:D}, as it must be.

The time scale $\gamma^{-1}$ for the dephasing does in fact not depend on $\gamma$ being large. The Lindblad equation \eqref{eq:Lindbladnm} always has eigenmodes with an eigenvalue $-\gamma$, namely all those matrices $\rho$ that are invariant under the translation $\rho_{n,m} \to \rho_{n+k,m+k}$ (assuming periodic boundary conditions) and have a vanishing trace. Translationally invariant deviations from the stationary state therefore relax with a rate $\gamma$. 

Since the Lindblad equation \eqref{eq:ULindblad} is linear in $\rho$ and has a simple form, a full analytical solution for the eigenmodes can be obtained by using spectral methods, see \cite{KenkreStochasticLiouville}.\\
\rnew{The diffusion equation \eqref{eq:diffnn} and the dephasing time scale $\gamma^{-1}$ hold for all unravellings of the Lindblad equation \eqref{eq:ULindblad} when the noise  term dominates over the Hamiltonian term. The implications for the wave function dynamics in the white-noise potential unravelling and in the quantum state diffusion unravelling is the subject of the following sections.}

\subsection{Broadening, subdiffusion and dephasing of a wave function in a white-noise potential}
\label{bsdwnp}
The dynamics \eqref{eq:WNPcoeff_old} of a wave function in a white-noise potential leads to a broadening of the wave function with time. The reason is that the fluctuating potential changes only the phase and not the amplitude at each lattice site, while the kinetic energy term by itself broadens the wave function. The diffusive broadening \eqref{eq:diffusion} of the second moment is true for all unravellings when averaged over an ensemble. But only in the white-noise potential unravelling the width of each wave function shows this broadening and increases as $\sigma(\hat x) \sim \sqrt t$.
This in turn means that  the position of the centre of mass of the wave function may move subdiffusively without violating relation \eqref{sq}.

Indeed, subdiffusive motion according to $\mathbf{M}[\langle \hat{x}\rangle^2] \sim \sqrt t$ was observed numerically  by \cite{Bouchaud1992,Saul1992}, and \cite{Saul1992} also gave a scaling argument. These authors studied the propagation of directed waves in random media, where the direction of propagation in the medium plays the role of the time dimension in our model. 

\begin{figure}
	\includegraphics[width=0.9\textwidth]{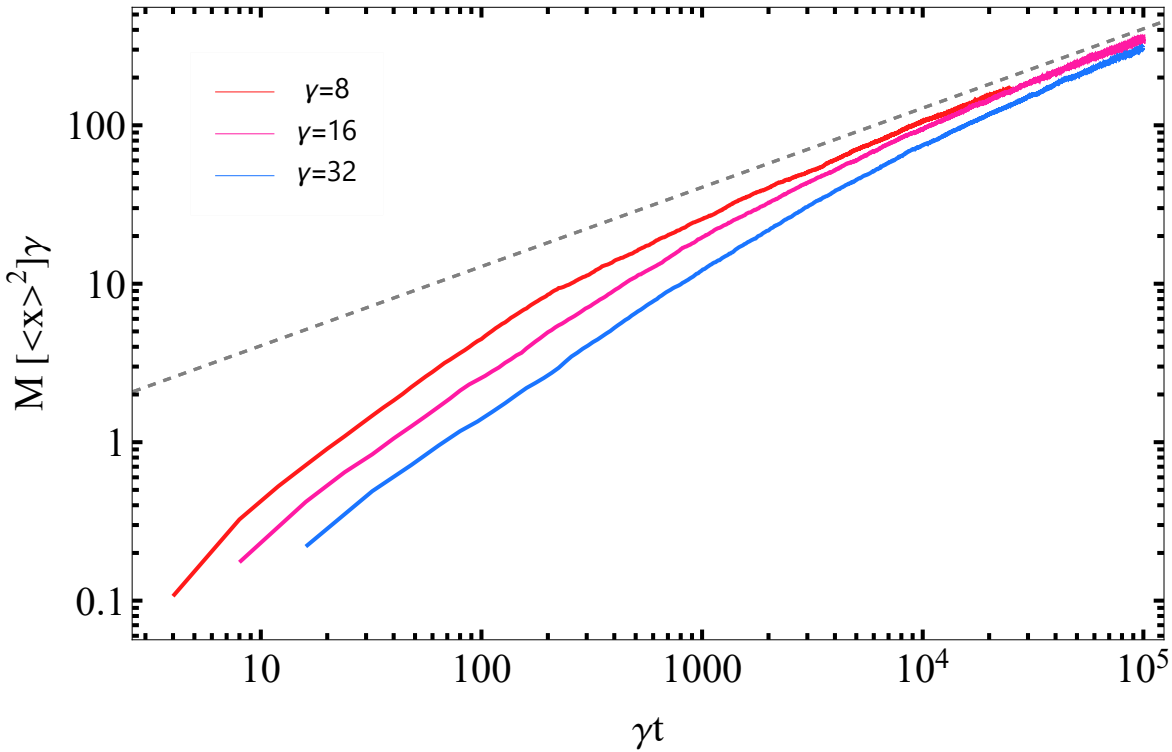}
	\caption{The ensemble average of the squared centre-of-mass position of the wave function in the white-noise potential model. The initial state was a gaussian wave packet with variance $\sigma^2=4$ and real phase at every lattice site. Here, the average was taken over $4\cdot 10^3$ trajectories. The solid grey line indicates a late time scaling $\propto t^{1/2}$.
}
	\label{fig:xsquared}
\end{figure}

Our simulations confirm this subdiffusive behavior, as shown in Figure \ref{fig:xsquared}.
Our intuitive explanation for the relation $\mathbf{M}[\langle \hat{x}\rangle^2]\sim \sqrt t$ is the following: 
We start from the fact that the width of the wave function increases as $\sqrt t$. This means that the weight of the wave function $\sum\limits_{n=n_0}^{n_0+l} |c_n|^2$ in a spatial section of length $l$ that is covered by the wave function decreases with time as $1/\sqrt t$. We take an $l$ that is not smaller than the spatial coherence length of the wave function. Between such sections, weight is shifted back and forth randomly. This means that each of the order of $\sqrt t$ additive contributions to the mean  $\langle \hat{x}\rangle$ changes by an amount proportional to $1/\sqrt t$ during a small time interval. The square $\langle \hat{x}\rangle^2 $ of the mean position therefore changes as $(\sqrt t)^2 / \sqrt t = \sqrt t$.  

Next, let us discuss the coherence time and coherence length of the wave function. The coherence time $\tau=\gamma^{-1}$ was calculated analytically in \cite{amir2009classical}. It can be read off equation \eqref{eq:WNPcoeff} immediately when the kinetic energy term is neglected. Then time appears on the right-hand side only in the combination $\gamma t$ (remember that $dW_n \propto \sqrt{dt}$), implying that the time scale is set by $\tau =\gamma^{-1}$. \\
The coherence length $\xi$ is closely connected to the coherence time and is the distance over which the values $c_m$ are affected by the values of other $c_n$ at earlier times. This coherence length must remain finite, even though the width of the wave function increases with time. Otherwise it would not be possible to construct unravellings that have a wave function width that stays finite.

We estimate the spatial coherence length in the limit of large $\gamma$ by starting from the coherence time $\tau =\gamma^{-1}$.
The distance over which a change in a $c_m$ affects during the time $\gamma^{-1}$ the values of neighbouring $c_n$ is proportional to $\gamma^{-1}$ since the kinetic energy contribution to $dc_n$ comes from free motion between lattice sites. 
We can back the estimation $\xi \propto \gamma^{-1}$ by taking together various results from the literature. In \cite{girvin1979exact}, the noise-averaged propagator of the model is calculated to be
\begin{equation}
\mathbf{M}\,G_{n}(\Delta t)\equiv \mathbf{M}\,[\bra{n} {\hat{U}(\Delta t) \ket{0}}]=e^{-\frac{\gamma}{2}\Delta t} g_n(\Delta t)
\end{equation}
with $\hat{U}=e^{-i\hat H t}$ being the time evolution operator, the index $n$ denoting the lattice sites, and $g_n(\Delta t)$ being the propagator in absence of noise. This propagator was calculated in \cite{Hammer} to be
\begin{equation}
    g_n(\Delta t) = i^nJ_n(2\Delta t)\, 
\end{equation}
with the $J_n$ being the Bessel functions. In the limit of small times $\Delta t \sim 1/\gamma$,
this becomes  $g_n \simeq i^n\frac{(\Delta t)^n}{n!}$. This decreases with distance $n$ as $1/\gamma^n$. For $n=1$, it is linear in $\Delta t$ for short times. We conclude from this that during the coherence time $\Delta t = \tau = \gamma^{-1}$ a change of the wave function at a site affects its nearest neighbours only with a weight $ 1/ \gamma$, so that the coherence length is also of the order $ 1/ \gamma$. This is in agreement with the finding above in section \ref{sec:timescaleseparation} that the elements in the $k$th off-diagonal of the density matrix are smaller by a factor $1/ \gamma^k$ than the diagonal elements.\\
\rnew{In conclusion, the white-noise potential leads to ever-spreading wave packets 
whose squared centre-of-mass position increases subdiffusively. 
Despite the spreading of the wave packet, there is a finite coherence time and a finite coherence length, both of which are proportional to   $1/\gamma$ in the limit of large noise.}

\subsection{Scaling of collapse rate and wave-packet width in the  quantum state diffusion unravelling}

The  quantum state diffusion unravelling highlights those features of the model that are less apparent in the white-noise potential unravelling. An initially broad wave function 'collapses' to a wave packet with a  width in the vicinity of the quantum mechanical coherence length. From then on, the wave packet performs a diffusive motion. 
The time scales of these processes can be quantified based on the results obtained in the previous sections. For large $\gamma$, the time scale of the initial collapse is set by $1/\gamma$, and it corresponds to the fast time scale of dephasing discussed above in Sec.~\ref{sec:timescaleseparation} for the density matrix. The diffusion of the wave packet on the lattice is characterized by the diffusion constant $D={4/\gamma}$, see \eqref{eq:D}, and it corresponds to the slow process mentioned in Sec.~\ref{sec:timescaleseparation}. This diffusive motion of the wave packet was shown above in Figure~\ref{fig:MSM}.

\begin{figure}
	\includegraphics[width=0.9\textwidth]{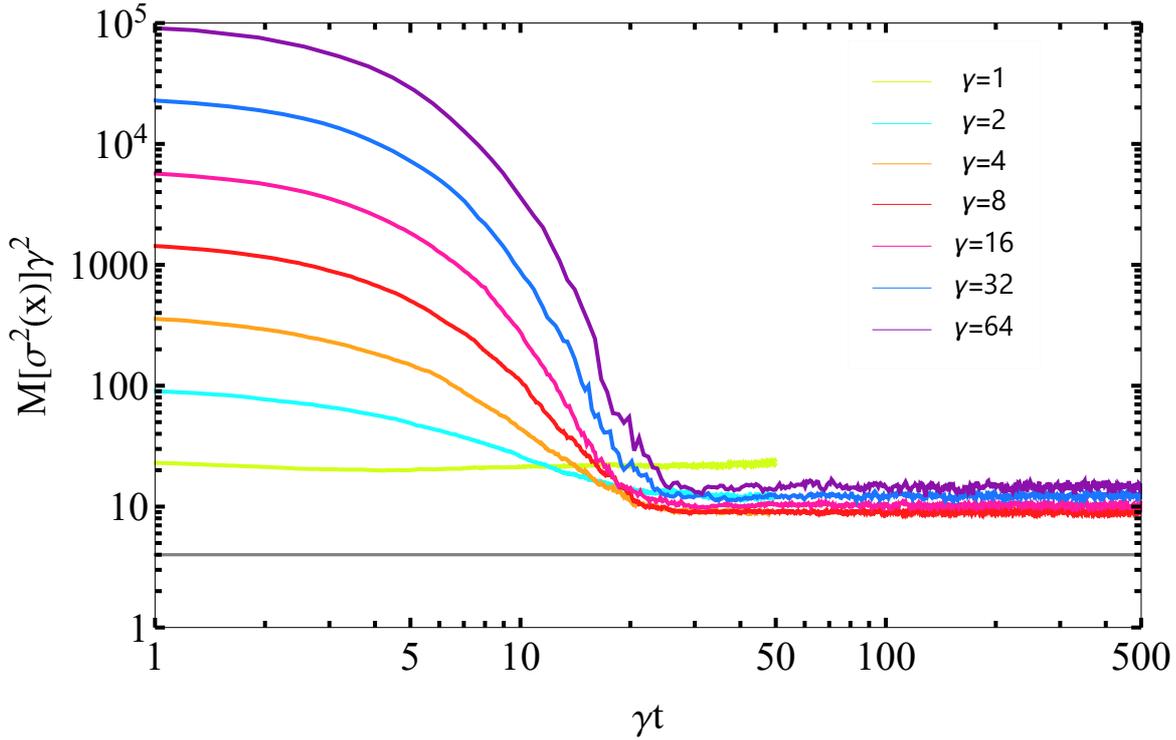}
	\caption{The ensemble-averaged quantum variance of the position operator in the  quantum state diffusion unravelling for different noise strengths $\gamma$, averaged over $5\cdot10^3$ trajectories for $\gamma=1,2,4$ and $10^4$ trajectories for $\gamma=8,16,32,64$, respectively. The initial state was a real-valued gaussian wave packet with variance $\sigma^2=25$. With the chosen scaling of the two axes, one can see that the collapse time scales as $1/\gamma$, while the asymptotic width decreases slower than $1/\gamma$. For comparison, the grey horizontal line gives the result obtained for the quantum jump model, see Eq.~\eqref{msjd}.
	}
	\label{fig:VarQSDLog}
\end{figure}

In the following, we will investigate how the width of the wave packets changes with time and with $\gamma$.
Figure~\ref{fig:VarQSDLog} shows the process by which the width of the wave function decreases to its stationary value. The collapse times become comparable when time is measured in units of $ 1 /\gamma$. If the width of the wave packet was identical to the phase coherence length calculated at the end of Section \ref{bsdwnp}, the variance multiplied by $\gamma^2$ should converge to an asymptotic value with increasing $\gamma$. However, the data indicate that the variance decreases slower than with $1/\gamma^2$. We checked this result by decreasing the time step of our numerical integration algorithm and by using real instead of complex noise. The asymptotic variance of the wave packet remained unchanged. For values $\gamma = 8$ and larger, this variance is so small that most of the weight of the wave function is concentrated on one site, with the two neighbours holding most of the remaining weight. In contrast to the quantum state diffusion unravelling, the time-averaged position variance obtained in the quantum jump unravelling decreases as $1/\gamma^2$, with the proportionality factor being 4, see Eq.~\eqref{msjd}. If the quantum state diffusion unravelling did lead to the same  decrease  $\sim 1/\gamma^2$ as the quantum jump unravelling, the curves in Figure~\ref{fig:VarQSDLog} would for large $\gamma$ converge to the black solid line.

\begin{figure}
\includegraphics[width=0.7\textwidth]{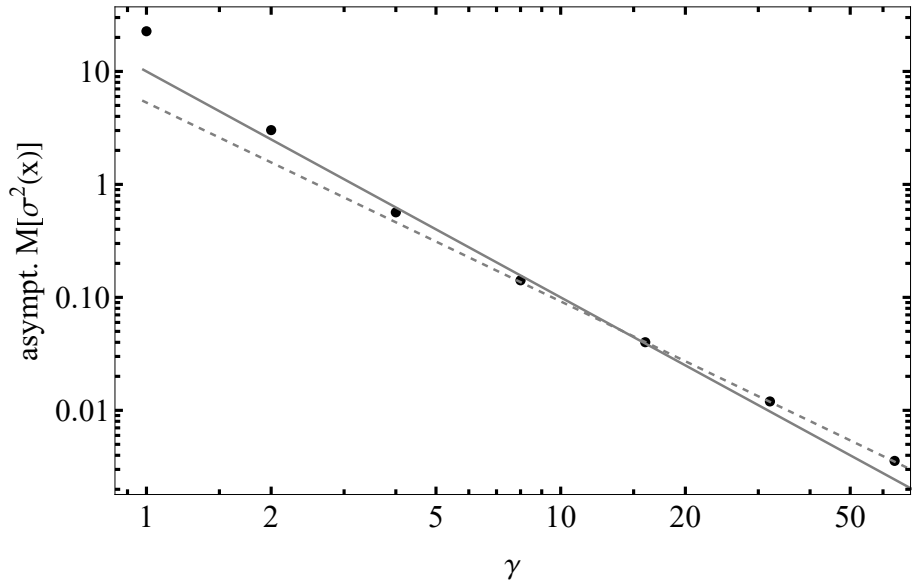}
\caption{Double-logarithmic plot of the time-averaged asymptotic variance of the data from Figure \ref{fig:VarQSDLog} for $\gamma t \geq 40$ versus the noise strength $\gamma$. The solid line is a power law $\propto \gamma^{-2}$, the dashed  line is a power law $\propto \gamma^{-1.76}$.}
\label{fig:VarQSDasympt}
\end{figure}

However, the decrease of the position variance wave packet with $\gamma$ in the quantum state diffusion unravelling is slower than $1/\gamma^2$.
Figure \ref{fig:VarQSDasympt} shows the asymptotic variance of the wave packet as function of $\gamma$, suggesting a decrease $\gamma^{-\kappa}$ with an exponent $\kappa$ in the range $1.7-1.8$.

\begin{figure}[h]
		\includegraphics[width=0.49\textwidth]{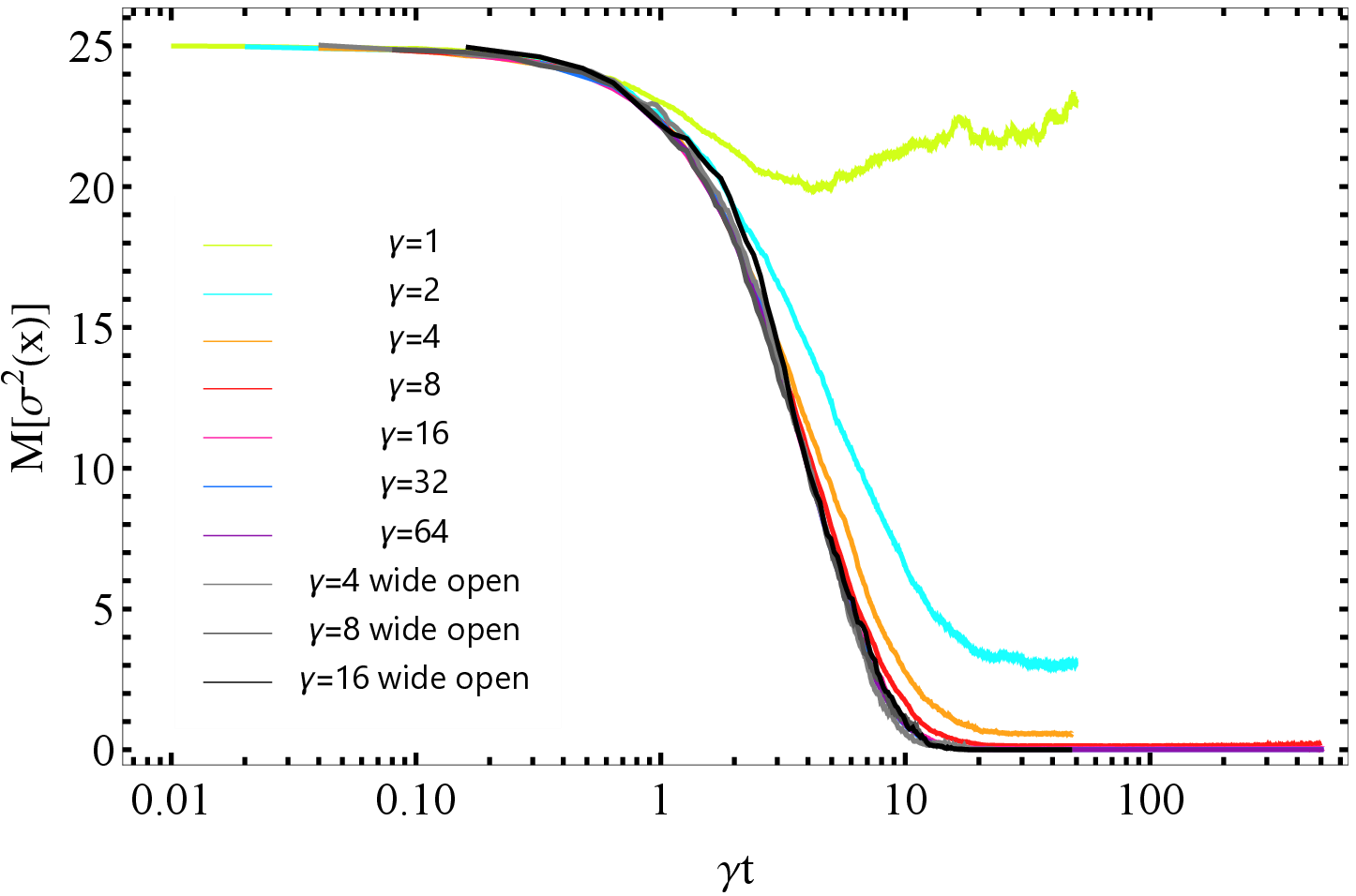}
	\includegraphics[width=0.49\textwidth]{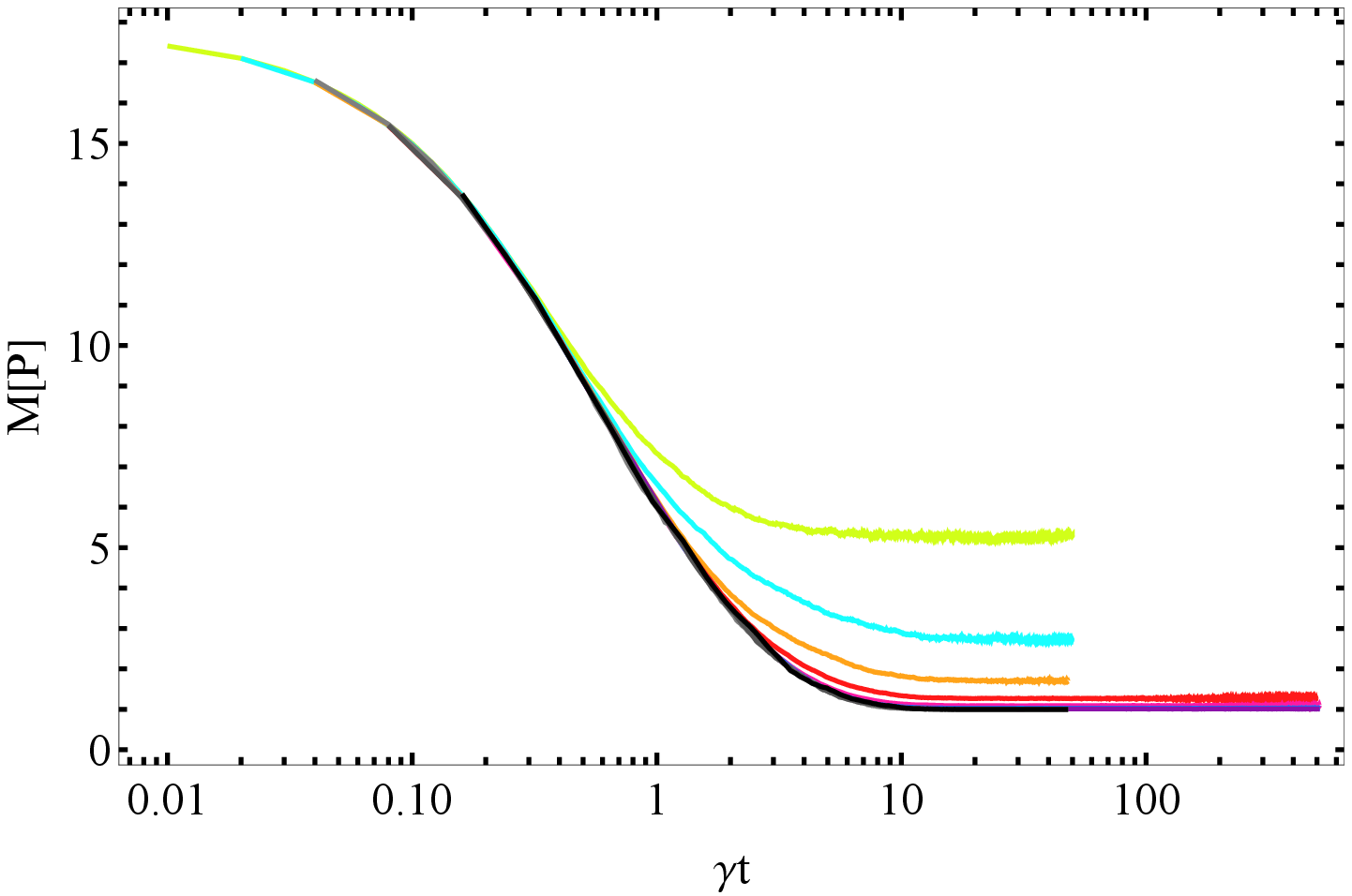}
	\caption{Ensemble-averaged quantum variance (left) and participation number (right) of the position operator in the quantum state diffusion unravelling with the full equation (colored) and for a wide-open system (gray scales), for different noise strengths $\gamma$, averaged over $5\cdot 10^3$ trajectories for $\gamma=1,2,4$ and $10^4$ trajectories for $\gamma=8,16,32,64$ and $10^3$ trajectories for the wide-open system, respectively. The initial state is a real-valued gaussian wave packet with variance $\sigma^2=25$.
}
	\label{fig:VarQSDVergleich}
\end{figure}

The limit of large $\gamma$ that we have considered is the limit  where the noise term dominates over the term with the Hamiltonian. In the literature \cite{Percival1998,percival1994localization}, such a limit is called that of a 'wide-open quantum system'. Often, the authors set $1/ \gamma=0$, i.e., they drop the term with the Hamiltonian altogether when studying wide-open quantum systems. Since it is not obvious that the limit $1/\gamma \to 0$ and the situation $1/\gamma = 0$ must be identical, we performed computer simulations also for the case where the Hamiltonian term is dropped altogether.
 Figure~\ref{fig:VarQSDVergleich} compares the two situations of small $1/ \gamma$ and vanishing $1/ \gamma$, showing that the data converge in the limit $1/\gamma \to 0 $ to those of the wide-open quantum system. The process of localization of the wave function can be understood by considering the dynamics of the weights $|c_n|^2$.  For the full quantum-state diffusion model, the weights $|c_n|^2$ evolve according to
\begin{align}
    d\vert c_n \vert^2  =& -2\text{Im}[c_n^\ast(c_{n+1}+c_{n-1})]dt \nonumber\\
      &+ 2\sqrt{\gamma} \vert c_n \vert ^2 (\text{Re}[d\xi_n] - \sum_k \vert c_k \vert^2 \text{Re}[d\xi_k])\, ,
\end{align}
with the first term being dropped when $1/\gamma$ is set to 0. The second term describes a multiplicative random walk, and the third term is a global coupling between all lattice sites that ensures that normalization is preserved.  Through the multiplicative noise, a few sites will accumulate most of the weight of the wave function if the kinetic energy term is dropped, until at the end one lattice site has all weight. By applying a calculation by Percival \cite{Percival1998} to our model, we prove this in the following.

From the time evolution \eqref{eq:UQSD} of the wave function, we can calculate the change of average position as 
\begin{align}
	  d\langle \hat{x} \rangle =&  \bra{\psi}\hat{x}\ket{d\psi} + \bra{d\psi}\hat{x}\ket{\psi} + \bra{d\psi}\hat{x}\ket{d\psi}\nonumber\\
	=& -i \langle [\hat{x},\hat{H}] \rangle dt
	+ 2\sqrt{\gamma}\sum_n (\langle \hat{x}  A_n \rangle - \langle \hat{x} \rangle \langle A_n \rangle ) \text{Re}[d\xi_n]\, ,
\end{align}
with the $A_n$ being the projection operators $\ket{n}\bra{n}$, as before. In the last step we have used the explicit expression $\hat{x}=\sum_n n\ket{n}\bra{n}$ for the position operator. Since the imaginary part of the noise changes only the phase of the wave function, only the real part occurs in this equation, and only the real part affects the rate of collapse. 
The mean change in the variance of the position is then
\begin{align}
\mathbf{M}\,[d \sigma^2(\hat{x})] =& \mathbf{M}\,[d\langle \hat{x}^2\rangle -2\langle \hat{x}\rangle d\langle \hat{x} \rangle - (d\langle \hat{x} \rangle )^2] \nonumber\\
=& -i\langle [\hat{x}^2,\hat{H}]\rangle dt
+ 2i \langle \hat{x} \rangle \langle [\hat{x},\hat{H}]\rangle dt \label{eq:dVarG} \\
&-  2\gamma\sum_n (\langle \hat{x}  A_n \rangle - \langle \hat{x} \rangle \langle A_n \rangle )^2 dt\, .\nonumber
\end{align}
For a wide-open quantum system, we keep only the last term and continue the calculation,
\begin{align}
\mathbf{M}\,[d \sigma^2(\hat{x})] 
=& - 2\gamma \sum_n (\langle \hat{x} {A_n} \rangle - \langle \hat{x} \rangle \langle A_n \rangle )^2 dt\nonumber\\
=& - 2\gamma\sum_n (n \vert c_n \vert^2 - \vert c_n \vert^2 \langle \hat{x} \rangle)^2 dt\nonumber\\
 =&- 2\gamma\sum_n  \vert c_n \vert^4 (n - \langle \hat{x} \rangle)^2dt.\label{eq:dVar}
\end{align}
Expression \eqref{eq:dVar} is always negative, unless the wave function is completely localized, i.e. $c_n=0$ for all but one $n$, for which $n=\langle \hat{x} \rangle$. 

This localization process is shown in Figure~\ref{fig:VarQSDVergleich}(a) for different values of $\gamma$. With increasing $\gamma$, the curves converge towards the curve obtained for a wide-open system when time is measured in units of $1/\gamma$.

Since distance plays no role in the uncoupled model where the Hamiltonian term has been dropped, a better measure than $\sigma^2$ of the increasing localization is the participation number. It is defined as 
\begin{equation}
    P= \frac 1 {\sum_n |c_n|^4}
\end{equation}
and is a measure of the number of sites on which a wave function sits. If the wave function is concentrated on one site, we have   
$P=1$, while $P=N$ if $N$ lattice sites all carry the weight $|c_n|^2=1/N$. Figure~\ref{fig:VarQSDVergleich}(b) shows the time evolution of $P$. Of course, all curves for the uncoupled system collapse on one curve.  For the coupled system, the curves approach an asymptotic curve for large $\gamma$, which is identical to the case  $1/\gamma=0$ of a wide-open quantum system, where the kinetic energy term is dropped altogether.

The time until the asymptotic participation number is reached is shorter than the time to reach the asymptotic variance since spatially separated non-vanishing parts of the wave function make a much larger contribution to the position variance than to the participation number.\\
\rnew{In contrast to the white-noise potential model, the wave functions in the quantum state diffusion model evolve towards narrow wave packets whose width is larger than the coherence length $1/\gamma$. 
The centre-of-mass of the wave packet moves diffusively with the diffusion constant \eqref{eq:D}. The scaling of the squared centre-of-mass position and the position variance for the different unravellings are summarized in Table \ref{tab:unravellings}.}

\begin{table}
\caption{Ensemble-averaged squared centre-of-mass position \textbf{M}\,[$\langle \hat{x} \rangle^2]$ and quantum variance of the position operator \textbf{M}$\,[\sigma^2(\hat{x})]$ for the three unravellings.}
\begin{tabular}{|c || c | c |}
 \hline
 Unravelling & \textbf{M}\,[$\langle \hat{x} \rangle^2]$ & \textbf{M}$\,[\sigma^2(\hat{x})]$ \\ [0.5ex] 
 \hline\hline
 white-noise potential & $\propto \sqrt{t}$ & increasing with time, $\propto t$ \\ 
 \hline
 quantum state diffusion & $\propto t$ & finite, $\propto \gamma^{-y}$ with $y< 2$\\
 \hline
 quantum jumps & $\propto t$ & finite, $\propto \gamma^{-2}$\\
 \hline
\end{tabular}
\label{tab:unravellings}
\end{table}

\section{Discussion}

In this paper, we have studied different mathematical representations of the same model, which is an electron in a one-dimensional lattice in the presence of thermal noise. Our starting representation of the model was a one-band tight-binding model with a spatially and temporally uncorrelated white-noise potential. In the ensemble description, the dynamics of the system is described by the Lindblad equation \eqref{eq:ULindblad} where the Lindblad operators are projections onto lattice sites. If the first term in \eqref{eq:ULindblad}, which is the kinetic energy term, is neglected due to the noise strength $\gamma$ being large, the stationary solutions of \eqref{eq:ULindblad} are the eigenfunctions of the projection operators, i.e., they are fully localized on one lattice site. Mixed ensembles composed of several of these localized solutions are also stationary solutions.  Such a result is often understood to mean that the thermal environment localizes electrons in space, in agreement with physical intuition and  the practice of solid-state theorists to describe electrons in room-temperature metals as spatially narrow wave packets.

However, when the kinetic-energy term is taken into account, no matter how small it is compared to the noise term, the only stationary solution is a constant distribution over all lattice sites, where all eigenstates of the projection operators have the same weight.
When noise is large compared to the kinetic-energy term, there is a time-scale separation between a fast decay of the phase coherence between lattice sites and a slow, diffusive motion of the wave function weights on the lattice. 

The same Lindblad equation could have been derived starting from a completely different model, namely a fully quantum mechanical model for the electron on a lattice interacting on each lattice site with a bath of macroscopically many degrees of freedom (so that their spectrum is continuous) that are uncorrelated between different lattice sites. There are several ways of obtaining an effective model for the electron only, in which the bath degrees of freedom are eliminated. The first one is a path integral approach that uses semiclassical approximations, similar to what was done in the famous Caldeira-Leggett model \cite{caldeira1983path}. Such a path integral approach has been performed by \cite{allinger1995non} to calculate the environmentally-induced loss of phase coherence in quantum transport problems. The second class of approaches starts from the combined density matrix of system and bath  and takes the trace over the bath degrees of freedom in order to obtain the time evolution of the reduced density matrix for the system alone. The generic result of such an approach is a Lindblad equation if suitable assumptions are made, namely  a short-term memory (justifying the Markov approximation), a rapid decay of reservoir correlations (implemented by using a continuum of modes), and a lack of back-action from the system on the bath (so that a product ansatz for their combined density matrix can be made) \cite{breuer2002theory}. 

The different possible ways of deriving the Lindblad equation of our system reflect opposed microscopic views of the thermal bath. In the fully quantum mechanical model, the loss of phase coherence of the electron wave function on the different lattice sites is due to decoherence, i.e. the entanglement of the electron with the surrounding degrees of freedom. If it was possible to control or reverse the time evolution of the bath, one would be able to see that the information about the initial state of the electron has not vanished altogether but has only dissipated into the environment. But neither a reversal nor control is possible for a true, physical heat bath, not even in principle. A true heat bath emits thermal radiation and is thus coupled to the rest of the world. Interacting with a heat bath is thus a truly irreversible process that increases the entropy of the universe. This means that the purely quantum mechanical picture behind the Lindblad equation cannot be distinguished empirically from the much simpler one chosen by us, which is a classical fluctuating potential. This fluctuating potential cannot induce decoherence in the narrow sense, but it has exactly the same effect on the wave function: phase coherence is destroyed and classical dynamics occurs beyond the quantum mechanical coherence length. The equivalence of the two descriptions hinges on the Lindblad operators being projectors on position eigenstates. 

Various authors \cite{schlosshauer2005decoherence,gu2019can} argue that decoherence and dephasing are fundamentally different things. However, their arguments apply only when a reversal or control of the environmental time evolution is possible in principle, as for instance in NMR experiments, where the nuclear spins are only weakly coupled to the phonons. 

When the Lindblad equation is unraveled in terms of a wave function that evolves on the lattice in a classical white-noise potential, the noise destroys the phase coherence between sites and the temporal phase correlation on each site. Although the width of the wave function increases as the square root of time, different parts of the wave function that are separated by more than the phase coherence length cannot show quantum interference. The reason is that the observation of an interference pattern requires a repeated preparation of the same state, which is then being measured. But since the noise cannot be controlled or reproduced, this is not possible for our system. The concept of a quantum wave function that extends over a distance that is larger than the coherence length is therefore not an empirically useful concept. The different parts of the wave function are independent from each other and can therefore be treated as each being present with a probability equal to its weight. 

The  quantum jump unravelling does not have these shortcomings as it gives rise to wave functions that have a width that agrees with the quantum mechanical coherence length. In
 this respect the quantum jump unravelling can be considered as the most 'physical' one: it describes a quantum particle by a wave function only over those distances over which the wave nature can be considered as an empirically testable concept. However, in order to arrive at this description, one has to take the unintuitive path from a wave function in a fluctuating potential through the Lindblad equation to a different unravelling of the Lindblad equation. Similar comments apply to the quantum-state diffusion unravelling, which is continuous in time and also leads to narrow wave packets, albeit not as narrow as the quantum mechanical coherence length. This unravelling has the awkward features that it is non-local and non-linear in the wave function. Our finding that the width of the wave packet decreases with inverse noise strength slower than the spatial coherence length is an additional reason to think that quantum state diffusion is not a good model for the 'real' wave function dynamics.

For a scientific realist, this freedom of describing a system in different ways that are conceptually of a very different nature but cannot be discriminated empirically, is deeply unsatisfying. We think that this ambiguity is due to the fact that the model is too simple to describe how a spatial collapse of the wave function occurs for a quantum particle in a heat bath. Neither the fully classical nor the fully quantum mechanical description of the environment can do the job. Since localization of a wave function is a non-linear phenomenon, non-linear equations are required if a spatially narrow wave function shall be the generic, robust outcome of a time evolution. The  quantum state diffusion model is non-linear, but an ensemble of such systems can still be written in terms of a Lindblad equation that is linear in the density matrix.

If the equivalence of the different unravellings shall be broken, different wave functions must be affected by the environment in different ways. A natural and plausible way how this can occur is that the environment itself responds to the wave function. 
This means that a feedback between wave function and environmental degrees of freedom should be taken into account. In fact, there exists a class of models that does exactly this, namely Discrete Nonlinear Schrödinger equations \cite{eilbeck2003discrete,kevrekidis2009discrete}.
They were constructed to model the interaction between electrons and phonons. The physical intuition on which such models are based involves elements from quantum and classical physics. But why not? In condensed matter theory, calculations that combine quantum and classical elements are widespread, and they are empirically adequate. For us, this is an indication that classical mechanics might be equally fundamental as quantum mechanics, with each of these two having their range of applicability.

\section*{Acknowledgements}

We thank Bernd Fernengel, Mathis Stumpf and Daniel Nevermann for helpful discussions. This research is part of the DFG-funded project DR300/16. \rnew{We acknowledge support by the Deutsche Forschungsgemeinschaft (DFG – German Research Foundation) and the Open Access Publishing Fund of Technical University of Darmstadt.}

\bibliography{bibarXiv.bib}
\bibliographystyle{unsrt}

\end{document}